\documentclass[twocolumn,aps,prl,showpacs]{revtex4}

\usepackage{amsmath}
\usepackage{graphicx}
\usepackage{dcolumn}
\usepackage{bm}

\begin{document}

\title{First Principle Simulations of Heavy Fermion Cerium Compounds Based
on the Kondo Lattice}
\author{Munehisa Matsumoto$^1$}
\author{Myung Joon Han$^{1}$}
\author{Junya Otsuki$^2$}
\author{Sergey Y. Savrasov$^{1}$}
\affiliation{$^{1}$ Department of Physics,University of California, Davis, California
95616, USA}
\affiliation{$^2$ Department of Physics, Tohoku University,
Sendai 980-8578, Japan}
\date{\today }

\begin{abstract}
We propose a new framework for first--principle calculations of
heavy--fermion materials. These are described in terms of the Kondo lattice
Hamiltonian with the parameters extracted from a realistic density
functional based calculation which is then solved using continuous--time
quantum Monte Carlo method and dynamical mean field theory. As an example,
we show our results for the N\'{e}el temperatures of Cerium--122 compounds
(CeX$_{2}$Si$_{2}$ with X=Ru, Rh, Pd, Cu, Ag, and Au) where the general
trend around the magnetic quantum critical point is successfully reproduced.
Our results are organized on a universal Doniach phase diagram in a
semi--quantitative way.
\end{abstract}


\maketitle

First--principle description of heavy--fermion materials has been a
challenging problem for a long time. The difficulty arises from the
dual nature of the electrons between localization and itinerancy due
to the large Coulomb repulsion energy $U$ at each site of the
lattice. Here the relatively well--localized f--electrons interact
with the itinerant s,p,d--electrons that form the conduction band. The
heavy--fermion systems are generally described as the Anderson
impurity model in the dilute limit~\cite{anderson_1961} or the
Anderson lattice model (ALM) in the dense limit, and the
first--principle description of% some of
them has been done by several
authors~\cite{han_1997, zoelfl_2001, held_2001, shim_2007}. With the
development of dynamical mean field theory (DMFT)~\cite{georges_1996}
and novel continuous--time quantum Monte Carlo (CT--QMC) solvers for
the impurity problem~\cite{rubtsov_2005,werner_2006, haule_2007}, the
ALM description has become quite successful except for a very low
temperature range. The numerically exact treatment of the Anderson
impurity problem is still very expensive if the temperature range of
$O(1)$ K is to be reached.  Thus, the first--principle description of
strongly--correlated materials around the quantum critical point
(QCP)~\cite{sachdev_1999}, which has recently been attracting a lot of
research interest is yet to be solved.

Here we attack the problem using the Kondo lattice model (KLM)%, or more
%general Coqblin--Schrieffer model~\cite{coqblin_1969}
trying to focus on the
low--energy physics of the ALM. We show how this new approach works for a
archetypical family of so called Cerium 122 compounds, CeX$_{2}$Si$_{2}$
(X=Ru, Rh, Pd, Cu, Ag, and Au), which has been one of the most extensively
studied strongly--correlated materials since the discovery of heavy--fermion
superconductor CeCu$_{2}$Si$_{2}$~\cite{steglich_1979}.
Strictly speaking, we deal with
the Coqblin--Schrieffer model~\cite{coqblin_1969}
with full $14$ fold degenerate f--shell but effectively the degeneracy
is lowered due to the spin--orbit and crystal--field splittings.
%which
%we fully take into account in our calculation.
With the localized
Kondo--impurity picture we can save the amount of the degrees of
freedom in our model by eliminating charge fluctuations, and we can
reach much lower temperature range as compared to the ALM
simulations. The conduction band in the model is given by the
hybridization function between the localized 4f orbitals and the
s,p,d--conduction bands calculated by the first--principle electronic
structure calculation based on the local--density approximation (LDA)
with Hubbard I~\cite{sergey} type of the self--energy for the f
electrons. Then the Kondo coupling is defined via the
Schrieffer--Wolff transformation~\cite{schrieffer_1966}, and the KLM
is solved with the new efficient CT--QMC Kondo impurity
solver~\cite{otsuki_2007} combined with DMFT.

Now we define the realistic KLM Hamiltonian for a given Cerium compound. The
general Coqblin--Schrieffer Hamiltonian is the following. 
\begin{equation}
\mathcal{H}=\sum_{k}\epsilon _{k}c_{k}^{\dag }c_{k}+J_{\mathrm{K}
}\sum_{i\alpha \alpha ^{\prime }}f_{i\alpha }^{\dag }f_{i\alpha
^{\prime }}c_{i\alpha ^{\prime }}^{\dag }c_{i\alpha }+\sum_{i\alpha
}\Delta^{\alpha}_{\rm splitting}f_{i\alpha }^{\dag }f_{i\alpha }, \label{KLM}
\end{equation}
Here $\epsilon _{k}$ is the conduction band, $J_{\mathrm{K}}$ is the
Kondo coupling, $\Delta^{\alpha}_{\rm splitting}$ is the crystal and spin--orbital
field, $c_{k}$ and $f_{i\alpha }$ are the annihilation operators for
the conduction and 4f electrons, respectively, with the orbital
$\alpha $ on the lattice site $i$.  To solve this Hamiltonian we first
need to define $J_{\mathrm{K}}$ and the conduction electron Green
function. For this we perform the first principle DFT calculation
within the local density approximation for s,p,d electrons plus the
Hubbard I approximation for the f electrons based on the
full--potential linearized muffin--tin orbitals (LMTO) method
\cite{sergey} and calculate the hybridization function\
\cite{han_1997} $\Im \Delta _{\alpha }(\epsilon )=\pi
\sum_{k}|V_{\alpha k}|^{2}\delta (\epsilon -\epsilon _{k}) \simeq
\pi|V_{\alpha}|^{2}\rho(\epsilon)$ where $V_{\alpha k}$ is the
hybridization matrix element and $\rho(\epsilon)$ is the density of states
of the conduction electrons at energy $\epsilon$ which we measure from the Fermi energy.
We use experimental lattice parameters for all materials that we study.

\begin{figure}[tbp]
\scalebox{0.9}{
\includegraphics{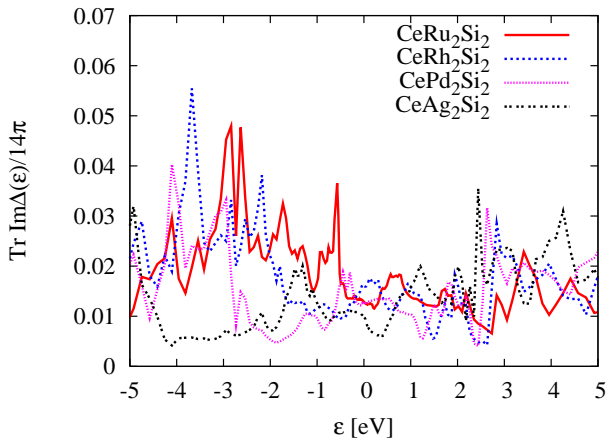}}
\caption{\label{all_hyb}(Color online)
The hybridization function $\mathrm{Tr}\Im
\Delta (\protect\epsilon )/14$ between the conduction band and the
$4f$--electrons calculated by LDA + Hubbard I for CeX$_{2}$Si$_{2}$
with X=Ru, Rh, Pd, and Ag. The origin of the energy is set to be the Fermi level.}
\end{figure}

The calculated
${\rm Tr}\Im \Delta(\epsilon)/(\pi N_{\rm F})\equiv [1/(\pi N_{\rm F})]\sum_{\alpha=1}^{N_{\rm F}}\Im \Delta_{\alpha}(\epsilon)$ is shown in Fig.~\ref{all_hyb}
for several representative CeX$_{2}$Si$_{2}$ materials with X=Ru, Rh,
Pd, and Ag. Here $N_{\mathrm{F}}=14$ is the total number of
degeneracy and the trace of $\Im \Delta$ is taken
over all of $N_{\mathrm{F}}$ states.
%Plotted on the vertical axis is the averaged $\Im \Delta
%$ per orbital, obtained dividing the trace over all of the orbitals by
%the number of orbitals $14$.
%We note that CeAg$_{2}$Si$_{2}$, which is
%an antiferromagnet that is far from the QCP, has a significantly
%weaker hybridization than the others.
We note that $\Im \Delta $
shows strong frequency dependence therefore in order to define
$J_{\mathrm{K}}$ an averaging over some frequency intervals needs to
be performed.

The Kondo coupling $J_{\mathrm{K}}$ is defined by the
Schrieffer--Wolff transformation\
\cite{schrieffer_1966,muehlschlegel_1968} as follows
\begin{equation}
J_{\mathrm{K}}=\frac{1}{\pi }\int_{-D_{\mathrm{cutoff}}}^{D_{\mathrm{cutoff}
}}d\,\epsilon \frac{\mathrm{Tr}\Im \Delta (\epsilon )}{N_{\mathrm{F}}}\left( 
\frac{1}{|\epsilon _{\mathrm{f}}|}+\frac{1}{(\epsilon _{\mathrm{f}}+U_{
\mathrm{eff}})}\right) .  \label{input_J_K}
\end{equation}
Here $\epsilon _{\mathrm{f}}$ is the location of the energy level of
4f orbital, and $U_{\mathrm{eff}}=U-J_{\mathrm{Hund}}$ is the
effective on--site Coulomb repulsion taking into account an effective
Hund coupling $J_{\mathrm{Hund}}$ that works in the virtual f$^{2}$
state. We set $\epsilon _{\mathrm{f}}=-2.5\mbox{[eV]}$ and
$U=5\mbox{[eV]}$ which is a typical value for Cerium compounds. The
Hund coupling $J_{\mathrm{Hund}}$ is explored around a realistic value
$1$ eV as is explained below.  In the present formulation,
$U_{\mathrm{eff}}$ incorporates all of the possible multiplet effects
in the virtual f$^{2}$ states and some systematic error comes in from
the setting of this value, but it is small enough to see the general
trend between the materials in the realistic Doniach phase diagram
that is obtained in Fig.~\ref{universal_Doniach_phase_diagram} in the
end. Here we have a band cutoff $D_{\mathrm{cutoff}}$ set to be $5$
[eV] which is large enough to make a universal description of the
low--energy physics~\cite{andrei_1983}.
%The number of the spin and
%orbital degeneracy is $N_{\mathrm{F}}=14$ and the trace of $\Im \Delta
%$ is taken over all of $N_{\mathrm{F}}$ states.

The portion of the conduction electron Green function $G_{\alpha }(\epsilon
) $ which has non--zero hybridization with the f--electrons is also
proportional to $\Delta _{\alpha }(\epsilon ).$ We define the normalized and
Hilbert--transformed $G_{\alpha }(i\omega )$ as follows
\begin{equation}
G_{\alpha }(i\omega )=\int_{-D_{\mathrm{cutoff}}}^{D_{\mathrm{cutoff}
}}d\,\epsilon \frac{\Im \Delta _{\alpha }(\epsilon )}{i\omega -\epsilon }
/\int_{-D_{\mathrm{cutoff}}}^{D_{\mathrm{cutoff}}}d\,\epsilon \Im \Delta
_{\alpha }(\epsilon ).  \label{input_Gc}
\end{equation}
The Eqs. (\ref{input_J_K}), (\ref{input_Gc}) provide necessary inputs
which are plugged into the CT--QMC and solved with DMFT
self--consistency loop.  The details of the CT--QMC algorithm for the
Coqblin--Schrieffer model are given in \cite{otsuki_2007}. These
definitions for the realistic model are designed in such a way that it
becomes exact in the limit of constant hybridization with the relevant
quantity $N_{\mathrm{F}}J_{\mathrm{K}}\rho (0)$ that determines the
behavior of the KLM \cite{otsuki_2007}.

The LDA results for $N_{\mathrm{F}}J_{\mathrm{K}}\rho (0)$ for the
target materials are given in Table\ \ref{material_table}. The level
splittings $\Delta _{\alpha }$ appeared in \ref{KLM} are implemented
as the difference of the positions of $\epsilon _{\mathrm{f}}$'s which
are used in the update probability as is described in
Ref.~\cite{otsuki_2007}. These level splittings are taken from the
literature and summarized in Table~\ref{material_table}. We checked
that our results for the N\'{e}el temperatures are robust against
small changes of factor of $O(1)$ on the level splittings. These
$\Delta _{\alpha }$ reduce the effective degeneracy close to
$N_{\mathrm{F}}=2$~\cite{note_N_F}. Thus we call our model
\textquotedblleft realistic Kondo\textquotedblright lattice instead of
the Coqblin--Schrieffer lattice even though we are actually doing the
multi--orbital model.

\begin{table*}
\caption{\label{material_table} Inputs (given by LDA and experiments
in the literature) and outputs for each material.  $\Delta_{\rm
spin-orbit}$ is spin-orbit splitting between $j=5/2$ and $j=7/2$
states.  Crystal field splitting of $j=5/2$ state produces three
doublets with energy $E_{0}<E_{1}<E_{2}$ with $E_{1}-E_{0}=\Delta_{\rm splitting}^{1}$
and $E_{2}-E_{0}=\Delta_{\rm splitting}^{2}$. $J_{\rm Hund} \sim$ 1 eV is used. The
$\Delta_{\rm splitting}^{1},\Delta_{\rm splitting}^{2},\Delta_{\rm splitting}^{\rm spin-orbit}$ are given in meV
unit and $T_{\rm N}$ in K.
Similar values for the level splittings were used in a recent work~\cite{vildosola_2005}.}
\begin{tabular}{cccccccc}
\hline
material & $N_{\rm F} J_{\rm K} \rho(0)$ & $\Delta_{\rm splitting}^{1}$ & $\Delta_{\rm splitting}^{2}$ & $\Delta_{\rm splitting}^{\rm spin-orbit}$ & $T_{\rm N}$(literature) &  $T_{\rm N}$(our results)\\ 
\hline
CeRu$_2$Si$_2$ & 0.144 & 19$^{a,b}$ & 34$^{a,b}$&$\sim 300$ & (paramagnetic)$^{h}$ & $ 0$\\ 
CeRh$_2$Si$_2$ & 0.180 & 26.7$^{c}$ & 58.7$^{c}$ & $\sim 300$ & $36$--$39$$^{h}$ & $\sim 70$ \\ 
CePd$_2$Si$_2$ & 0.140 & 19$^{d,e}$ & 24$^{d,e}$ & 230$^{d}$ & $10$$^{h}$ & $\sim 50$ \\ 
\hline
CeCu$_2$Si$_2$ & 0.146 & 32$^{b,f,g}$ & 37$^{b,f,g}$  & $\sim 300$ & (paramagnetic)$^{h}$ &  $\sim 70$ \\ 
CeAg$_2$Si$_2$ & 0.110 & 8.8$^{e}$ & 18.0$^{e}$ & $\sim 300$ & $8$--$10$$^{h}$ & $\sim 30$\\ 
CeAu$_2$Si$_2$ & 0.125 & 16.5$^{e}$ & 20.9$^{e}$ & $\sim 300$ & $8$--$10$$^{h}$ & $\sim 50$\\ 
\hline
\end{tabular}\\
$^a$ Ref.~\cite{zwicknagl_1992}
$^b$ Ref.~\cite{ehm_2007}
$^c$ Ref.~\cite{settai_1997}
$^d${Ref.~\cite{hansmann_2008}}
$^e${Ref.~\cite{severing_1989}}
$^f${Ref.~\cite{goremychkin_1993}}
$^g${Ref.~\cite{note_crystal_field_CeCu2Si2}}
$^h${Ref.~\cite{endstra_1993}}\\
%$^i${caused by the closeness to the QCP as can be seen in Fig.~\protect\ref{material_specific_Doniach_phase_diagram}.}
\end{table*}

\begin{figure}[tbp]
\includegraphics[width=8cm]{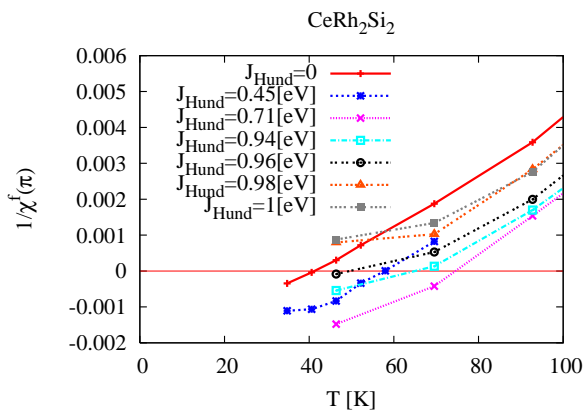}
\caption{\label{CeRh2Si2_analyses_for_Doniach_phase_diagram}
(Color online)
Determination of N\'{e}el temperatures for several settings
of the Hund coupling $J_{\mathrm{Hund}}$. We see that the N\'{e}el
temperature disappears at the point where $J_{\mathrm{Hund}}$ takes
the value between $0.96$ and $0.98$ eV. }
\end{figure}

We apply the above framework for the KLM description of
CeX$_{2}$Si$_{2}$ with X=Ru, Rh, Pd, Cu, Au, Ag.  We do the following
analyses with several settings of $U_{\mathrm{eff}}=U-J_{
\mathrm{Hund}}$ for $0\leq J_{\mathrm{Hund}}\overset{<}{\sim }1\ $eV
for each of the material. For a given material and given parameter
set, we determine the N\'{e}el temperature by looking at the
temperature dependence of staggered susceptibility and locating at
which temperature it diverges.  Here we follow the formalism of DMFT
for the localized f--electron systems as given
in~\cite{otsuki_2009_formalism} and use the same method as was
utilized for model calculations
in~\cite{otsuki_2009_Doniach}. Regarding the realistic input of the
Green's function as is depicted in Fig.~\ref{all_hyb} , we make an
approximation in the calculation of the staggered magnetic
susceptibility for the 4f--electrons by decoupling the two--particle
density of states $\rho (\epsilon _{1},\epsilon _{2})=\delta (\epsilon
_{1}+\epsilon _{2})\rho (\epsilon _{1})$ as if there is a nesting
property which becomes exact when the 4f--electrons are on a
hypercubic lattice. Thus the tendency to the antiferromagnetic order
would be overestimated in addition to having the infinite--dimensional
nature in the DMFT solution to the lattice problem.  The data specific
to CeRh$_{2}$Si$_{2}$ with which we determine the N\'{e}el
temperatures for several settings of $J_{\mathrm{Hund}}$ are shown in
Fig.~\ref{CeRh2Si2_analyses_for_Doniach_phase_diagram}. In this way
for each of the material we look at the magnetic phase transitions for
several $J_{\mathrm{K}}$'s by varying corresponding
$J_{\mathrm{Hund}}$'s.

\begin{figure}[tbp]
\begin{tabular}{l}
(a) \\
\scalebox{0.85}{\includegraphics{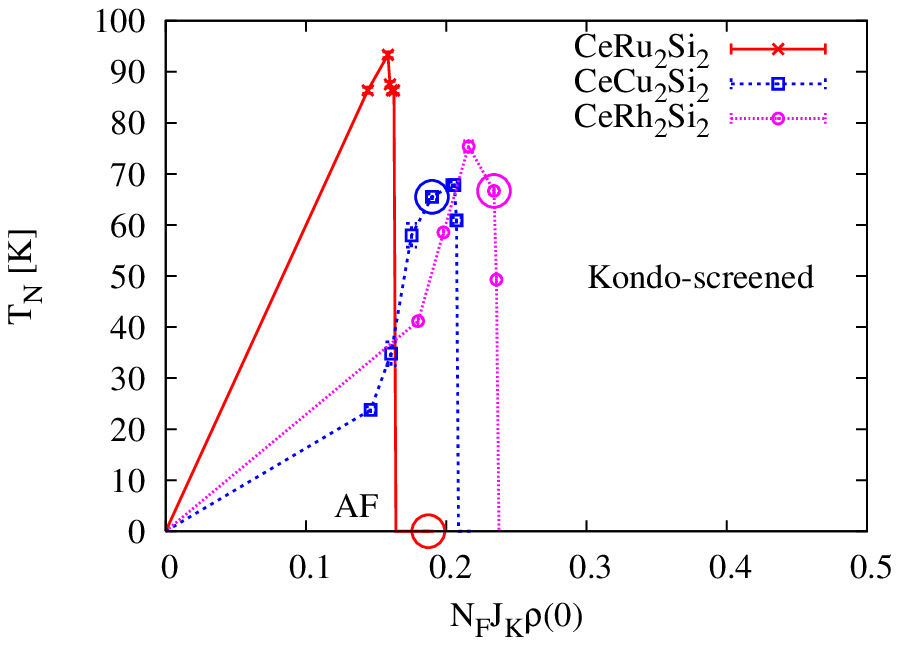}}\\
(b) \\
\scalebox{0.85}{\includegraphics{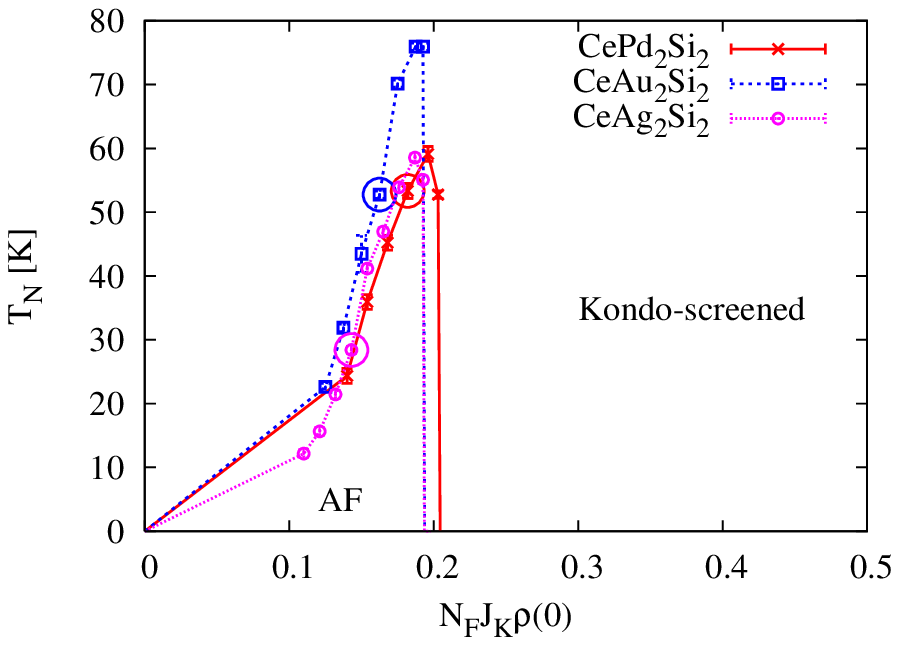}}
\end{tabular}
\caption{(Color online)
Material specific Doniach phase diagrams for
CeX$_{2}$Si$_{2}$ with (a) X=Ru,Cu, and Rh and with (b) X=Pd, Au, and
Ag.
%The left--most data point for each of the material corresponds to
%the data with $J_{\mathrm{Hund}}=0$ and
The realistic results with
$J_{\mathrm{Hund}}=0.94$ eV are marked with the larger plot symbols. }
\label{material_specific_Doniach_phase_diagram}
\end{figure}

\begin{figure}[tbp]
\scalebox{0.9}{\includegraphics{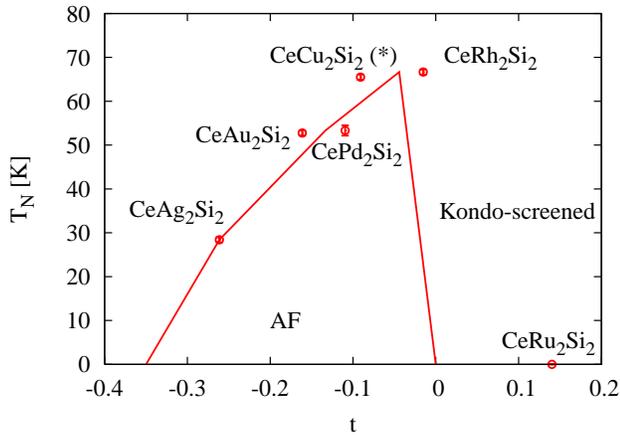}}
\caption{(Color online)
Universal Doniach phase diagram for the material family of
CeX$_{2}$ Si$_{2}$. The horizontal axis is defined as follows:
$t\equiv \lbrack N_{\mathrm{F}}J_{\mathrm{K}}\protect\rho
(0)-N_{\mathrm{F}}J_{\mathrm{K}} \protect\rho
(0)|_{\mathrm{QCP}}]/N_{\mathrm{F}}J_{\mathrm{K}}\protect\rho(0)|_{\mathrm{QCP}}$. The
line is guide for the eye. We plot CeCu$_{2}$Si$_{2}$ with an asterisk
mark to note as it apparently has a finite N\'{e}el temperature but
that just reflects the result that CeCu$_{2}$Si$_{2}$ is very close to
QCP as can be seen in
Fig.~\protect\ref{material_specific_Doniach_phase_diagram}. }
\label{universal_Doniach_phase_diagram}
\end{figure}

As was first discussed by Doniach~\cite{doniach_1977,doniach_1987} and
subsequently by many authors, Kondo lattices have two representative
energy scales, namely the magnetic ordering energy that is proportional
to $(J_{\mathrm{K}}/N_{\mathrm{F}})^{2}\rho(0)$ and the Kondo
screening energy which behaves like $\exp \left(
-1/N_{\mathrm{F}}J_{\mathrm{K}}\rho (0)\right) $.  For small
$J_{\mathrm{K}}$'s the former wins but as $J_{\mathrm{K}}$ becomes
larger the exponential growth of the latter dominates at some
point. Thus a given system can realize in either magnetically ordered
phase or non--magnetic Kondo--screened phase. Between these two phases
at zero temperature there is thought to be a QCP.  We explore this
Doniach phase diagram for each material and find the
material--specific QCP.  We take the data with the setting
$J_{\mathrm{Hund}}=0.94$ eV as our realistic result for each material
as this strength of the Hund coupling is close to the realistic value
and also gives reasonable trend over all materials in the family. Thus
obtained Doniach phase diagrams for all of the materials are shown in
Fig.~\ref{material_specific_Doniach_phase_diagram}.

Now we can plot all of the six materials CeX$_{2}$Si$_{2}$
(X=Ru,Rh,Pd,Cu,Ag,Au) on the universal Doniach phase diagram in the
same spirit as was done by Endstra et al. in 1993~\cite{endstra_1993}
but with a different horizontal axis. In the material--specific
Doniach phase diagrams in
Fig.~\ref{material_specific_Doniach_phase_diagram}, we see that the
locations of QCP's on the $N_{\mathrm{F}}J_{\mathrm{K}}\rho (0)$ are
not actually universal~\cite{yi-feng_2008}. So we measure the distance
between the QCP and the material's realistic location on the
horizontal axis, $N_{\mathrm{F}}J_{\mathrm{K}}\rho
(0)-N_{\mathrm{F}}J_{\mathrm{K}}\rho (0)|_{\mathrm{QCP}}$, and plot
the N\'{e}el temperatures with respect to the value of the distance to
QCP defined as $t\equiv \lbrack N_{\mathrm{F}}J_{\mathrm{K}}\rho
(0)-N_{\mathrm{F}}J_{\mathrm{K}}\rho
(0)|_{\mathrm{QCP}}]/N_{\mathrm{F}}J_{\mathrm{K}}\rho
(0)|_{\mathrm{QCP}}$. The result is shown in
Fig.~\ref{universal_Doniach_phase_diagram}. We expect a systematic
error bar in the estimation of the value on the horizontal axis
especially around the QCP but these possible
systematic errors are small enough to discern the locations of
CeX$_{2}$Si$_{2}$ (X=Ag,Au) and CeRh$_{2}$Si$_{2}$.  The
antiferromagnet CeRh$_{2}$Si$_{2}$ and the paramagnet
CeCu$_{2}$Si$_{2}$ are mixed up within the present level of resolution
which is manifested by the result that finite N\'{e}el temperature is
plotted for CeCu$_{2}$Si$_{2}$ . However, this is actually caused by
the proximity of this material to its QCP as can be seen from
Fig.~\ref{material_specific_Doniach_phase_diagram} . So our
numerical result is consistent with the experimental result that
CeCu$_{2}$Si$_{2}$ is a superconductor at ambient pressure and is
thought to be close to the QCP.

We note that the valence fluctuations which we ignored in our
simulation could be important in the realization of the non--magnetic
ground state.
Indeed it is known that there are some valence
fluctuations in CeCu$_2$Si$_2$~\cite{ansari_1987} and CeRu$_{2}$Si$_{2}$~\cite{yano_2008}.
This might make the possible systematic error relatively
larger on the right-hand side of our phase diagram~\cite{holmes_2004,otsuki_2009_large_fs}.
Nevertheless at the present level of description we believe
that the realistic KLM works
%with a certain \textquotedblleft
%renormalized\textquotedblright\ parameter set for the KLM to
%incorporate the effects of valence fluctuations
because the number of 4f electrons in Cerium ion is still very close to one~\cite{ansari_1987,yano_2008}.
At least for the impurity problem
the convergence to the Kondo impurity picture in
large $|\epsilon _{\mathrm{f}}|/(\rho (0)V^{2})$ limit of the Anderson
model was discussed exactly~\cite{schlottmann_1984}.
Careful comparison between the Anderson lattice
and the Kondo lattice regarding the valence fluctuation issue is
interesting, especially for CeCu$_{2}$Si$_{2}$, and further work is
ongoing in this direction.

MM thanks Y.--F. Yang, P. Werner, and H. Shishido for helpful
discussions and N. Matsumoto for continuous supports. Discussions with
the participants in ICAM-DCHEM workshop in August 2008 are
acknowledged. This research is supported by NSF Grant No. DMR-0606498
and by DOE SciDAC Grant No. SE-FC02-06ER25793.  Numerical computations
are performed using TeraGrid supercomputer grant No.  090064.


\begin{thebibliography}{99}
\bibitem{andrei_1983} The definition of $J_{\mathrm{K}}$ should always be
combined with the cutoff scheme for the conduction band as is discussed in
N. Andrei \textit{et al.}, Rev. Mod. Phys. \textbf{55}, 331 (1983).
\end{thebibliography}
\end{document}